\documentclass[10pt]{iopart}

\newcommand{\eqn}[1]{(\ref{#1})}

\def\be{\begin{equation}}
\def\te{\end{equation}}
\def\ee{\end{equation}}
\def\ba{\begin{eqnarray}}
\def\bea{\begin{eqnarray}}
\def\nn{\nonumber\\}
\def\tea{\end{eqnarray}}
\def\ea{\end{eqnarray}}
\def\eea{\end{eqnarray}}

\begin{document}

\title[Collapse times for attractive BEC]{Collapse times for attractive Bose-Einstein condensates}

\author{Esteban Calzetta}

\address{Physics Department, FCEN-UBA and CONICET - Argentina}
\ead{calzetta@df.uba.ar}
\begin{abstract}
We argue that the main mechanism for condensate collapse in an attractive Bose-Einstein condensate is the loss of coherence between atoms a finite distance apart, rather than the growth of the occupation number in non condensate modes. Since the former mechanism is faster than the latter by a factor of approximately $3/2$, this helps to dispel the apparent failure of field theoretical models in predicting the collapse time of the condensate.
\end{abstract}

\pacs{03.75.Kk, 03.75.Nt, 03.70.+k}
\noindent{\it Keywords}: BEC, condensate collapse
\maketitle


\section{Introduction}
The so-called \textit{Bose Nova} experiment on the collapse of a Bose-Einstein condensate with attractive interactions \cite{JILA01,Cla03,CoThWi06} has opened up a fascinating window in the far out of equilibrium behavior of these systems. The experiment has been analyzed from a number of perspectives \cite{GFT01,YUR02,SS02,CH03,SU03,BJM04,ADH04} and it is fair to say that we have a good qualitative understanding of the phenomenon. However, at the quantitative level certain anomalies persist.

In this paper we shall deal with the apparent failure of existing models in predicting the collapse time scale $t_c$ for the condensate, in the regime where the scattering length $a$ is just below the critical value $-a_c$. In \cite{CH03} the scaling law 

\begin{equation}
t_c\propto\left[\frac{\left|a\right|}{a_c}-1\right]^{-1/2}
\end{equation}
is proposed, which fits well the experimental results. However, the proportionality constant is not derived. The authors of \cite{CH03} claimed that the proper proportionality constant could be derived from a complete field theoretic calculation, but when the calculation was actually done \cite{SRH03,WHS04,WDBDBHS06}, it failed to produce a satisfactory prediction.

In this paper we shall present a qualitative analysis of the collapse time for a condensate trapped in a flat box \cite{YUR02} with periodic boundary conditions. Unlike previous analysis, we shall assume that the total number of particles in the condensate remains fixed \cite{GirArn59,GirArn98,Gar97,CasDum98,GarMor06}. Under these constraint, the condensate occupation number is properly defined as the greatest eigenvalue of the one particle density matrix (to be defined below)\cite{PO56}. Given the assumed geometry, the corresponding eigenmode is necessarily homogeneous, so the eigenvalue is just the integral of the one particle density matrix with one argument fixed, and the other ranging over the confining box.

Since the overall normalization of the one particle density matrix is determined by the overall density of the gas (see below), the  most important factor in the evolution of the condensate occupation number is how fast the density matrix falls off, measuring the degree of coherence among atoms at finite distances. We shall argue below that the one particle density matrix is approximately Gaussian with a variance which decays in time as $\mathrm{exp}\left\{-\gamma t\right\}$, with 

\begin{equation}
\gamma =\tau^{-1}\sqrt{\frac{\left|a\right|}{a_c}-1}\label{gamma}
\end{equation}
where

\begin{equation}
\tau^{-1}=\frac{2\pi^2\hbar}{ML^2}\label{tau}
\end{equation}
is the frequency of the first excited states for a particle in the box; here $M$ is the mass of an atom and $L$ is the size of the box. Therefore, after integrating over the three dimensional box we obtain that the condensate occupation number decays as $\mathrm{exp}\left\{-3\gamma t\right\}$. 

The expectation number in the first excited state, as computed from the Gross - Pitaievskii equation, the Hartree - Fock - Bogoliubov or the Popov approximation would grow only as $\mathrm{exp}\left\{2\gamma t\right\}$. Therefore,  condensate collapse from loss of coherence between atoms is faster than the estimate from loss of particles to excited modes by a factor $3/2$. For comparison, note that a detailed calculation of the collapse time for $a=-10a_c$ yields a predicted value of $10\mathrm{ms}$ against an experimental value of $\left(6\pm 1\right)\mathrm{ms}$ \cite{WDBDBHS06}. Therefore a factor of three halves goes a long way to solve the existing puzzle.

This paper is organized as follows. In next section we present the model of a cold trapped Bose gas, introduce a suitable set of density and phase variables and solve the Heisenberg equations in the linearized approximation. In the following section, we apply these results to derive the evolution of the condensate particle number and thereby our main result. In Section 4 we compare this results to the particle number conserving, Hartree-Fock Bogoliubov and Popov approaches. We close the paper with some brief final remarks, and give supplementary technical details in the Appendix.

\section{The model}
The idea is to analyze the Bose Nova experiment with the tools we have developed  to handle the Mott transition in \cite{CHR06}. The starting point is a second -
quantized field operator $\hat{\Psi} \left( \mathbf{x},t\right) $ which removes an
atom at the location $\mathbf{x}$ at times $t$. It obeys the canonical
commutation relations

\begin{equation}
\left[ \hat{\Psi} \left( \mathbf{x},t\right) ,\hat{\Psi} \left( \mathbf{y},t\right) \right] =0
\end{equation}
\begin{equation}
\left[ \hat{\Psi} \left( \mathbf{x},t\right) ,\hat{\Psi} ^{\dagger }\left( \mathbf{y},t\right) \right]
=\delta \left( \mathbf{x-y}\right)  \label{etcr}
\end{equation}
The dynamics of this field is given by the Heisenberg equations of motion

\begin{equation}
-i\hbar \frac{\partial }{\partial t}\hat{\Psi} =\left[ \hat{\mathbf{H}},\hat{\Psi} \right]
\end{equation}
where $\hat{\mathbf{H}}$ is the Hamiltonian. The theory is invariant under a
global phase change of the field operator
\begin{equation}
\hat{\Psi} \rightarrow e^{i\theta }\hat{\Psi} ,\qquad \hat{\Psi} ^{\dagger }\rightarrow
e^{-i\theta }\hat{\Psi} ^{\dagger }  \label{global}
\end{equation}
The constant of motion associated with this invariance through
Noether theorem is the total particle
number.

To progress further, we need a specific model for the atom-atom interactions.
In principle, we should specify the atom-atom interaction potential. However, in
many applications it is enough to know the cross section $\sigma$
for low energy spherically symmetric scattering of two identical bosons. We introduce
the scattering length $a$ through $\sigma\equiv 8\pi a^{2}$, where the factor $8\pi$ involves both integration over all scattering angles and Bose enhancement factors. We shall adopt as model atom-atom interaction
a contact potential $U\delta \left(\mathbf{x}\right)$. This is expected to be a good approximation as long as the distance between atoms is much greater than both the scattering length and the distance out to which the fundamental atom-atom interaction is important \cite{PESM02}. To reproduce the right scattering length we need $U=4\pi \hbar ^2a/M$, where $M$ is the mass of the atoms. A positive value of $a$ means a repulsive interaction; we adopt the convention that an attractive interaction is described by a negative value of $a$.

Assuming a contact atom-atom potential we get the Hamiltonian

\begin{equation}
\hat{\mathbf{H}}=\int d\mathbf{x}\;\left\{ \hat{\Psi} ^{\dagger }\hat{H}\hat{\Psi} +\frac {U}2\hat{\Psi} ^{\dagger
2}\hat{\Psi} ^2\right\}  \label{nbodyh}
\end{equation}
The
single-particle Hamiltonian $\hat{H}$ is given by

\begin{equation}
\hat{H}\hat{\Psi} =-\frac{\hbar ^{2}}{2M}\nabla ^{2}\hat{\Psi} +V_{trap}\left( \mathbf{x}\right) \hat{\Psi}
\label{sparth}
\end{equation}
$V_{trap}$ denotes a confining trap potential. Then the
Heisenberg equation of motion

\begin{equation}
i\hbar \frac{\partial }{\partial t}\hat{\Psi} =\hat{H}\hat{\Psi} +U\hat{\Psi} ^{\dagger }\hat{\Psi} ^{2}
\label{Heisenberg}
\end{equation}
is also the classical equation of motion derived from the action

\begin{equation}
S=\int dtd\mathbf{x}\;i\hbar \Psi ^{*}\frac{\partial }{\partial t}\Psi -\int dt\;%
\mathbf{H}  \label{action}
\end{equation}
placing hats everywhere. For simplicity we shall replace the trap potential by a flat bounding box of volume $V=L^{3}$ with periodic boundary conditions. Yurovsky has demonstrated that this is enough for a qualitative treatment of the \textsl{Bose Nova} \cite{YUR02}. We also assume that we have a finite total number of particles $N$, which remains fixed through the evolution (that is, there is no particle loss to the environment).

\subsection{Density and phase variables in the CTP formulation}

To analyze further this model we shall adopt density-phase variables \cite{MAD27,HAL81}. These variables have been extensively used to study dynamical problems, including the Mott transition \cite{DSZB04}. 
This will set the stage for a further canonical transformation to a more
convenient set of degrees of freedom, to be carried out in the next Section.

In the path integral representation, quantum amplitudes are given in terms of functional integrals over complex fields $\Psi$ and $\Psi^{\dagger}$ associated to the destruction and creation operators. Our starting point is the Madelung representation \cite{MAD27,HAL81}

\begin{equation}
\Psi \left( \mathbf{x},t\right)=\left[ \exp -i\varphi \left( \mathbf{x},t\right)\right] \sqrt{\rho\left( \mathbf{x},t\right)}  \label{mad1}
\end{equation}

\begin{equation}
\Psi^{\dagger }\left( \mathbf{x},t\right)=\sqrt{\rho\left( \mathbf{x},t\right)}\left[ \exp i\varphi \left( \mathbf{x},t\right)\right].  \label{mad2}
\end{equation} 

In the canonical formalism, the fields $\rho$ and $\varphi$ become operators with commutation relations

\begin{equation}
\left[ \hat{\rho}\left( \mathbf{x},t\right),\hat{\varphi} \left( \mathbf{y},t\right)\right] =-i\delta \left( \mathbf{x-y}\right),
\end{equation}
Within the path integral we allow the phases $\varphi$ to take all real values, and therefore so do the conjugated density operators $\rho $ \cite{CHR06,Kli90}. This makes the square roots in \eqn{mad1} and \eqn{mad2} problematic. It is best to adopt a new set of variables where square roots do not appear, as we shall do presently. For further discussion of density-phase variables in continuum theories see \cite{CAS04}. 

We adopt the formalism developed in \cite{CHR06} to describe the transition from the superfluid to the Mott insulator state in an optical lattice. To compute expectation values, we shall use the closed time-path formalism, where we choose the independent variables as follows. 
In the first branch, we define a new
(complex) variable $\chi^{1} \left( \mathbf{x},t\right)$ from

\begin{equation}
\Psi^{1} \left( \mathbf{x},t\right)=\exp
\left[ -i\chi^{1}\left( \mathbf{x},t\right)\right]\label{represen1}
\end{equation}

\begin{equation}
\Psi^{1\dagger } \left( \mathbf{x},t\right)=\rho^{1}\left( \mathbf{x},t\right)\;\exp \left[ i\chi^{1}\left( \mathbf{x},t\right)\right]
\end{equation}
On the second branch we write instead

\begin{equation}
\Psi^{2\dagger} \left( \mathbf{x},t\right)=\exp
\left[ i\chi^{2\dagger}\left( \mathbf{x},t\right)\right]
\end{equation}

\begin{equation}
\Psi^{2 } \left( \mathbf{x},t\right)=\exp \left[ -i\chi^{2\dagger}\left( \mathbf{x},t\right)\right]\;\rho^{2}\left( \mathbf{x},t\right)
\end{equation}
In the canonical formulation, the fields $\chi$ and $\rho$ become operators with commutation relations \cite{CHR06}

\begin{equation}
\left[ \hat{\rho}\left( \mathbf{x},t\right),\hat{\chi} \left( \mathbf{y},t\right)\right] =-i\delta \left( \mathbf{x-y}\right),
\end{equation}
The dynamics of these operators is given by the Hamiltonian

\begin{equation}
\hat{H}\left( \hat{\rho},\hat{\chi} \right) =\int d\mathbf{x}\left\{\frac{\hbar^2}{2M}\left(\hat{\rho}\nabla\hat{\chi}-i\nabla\hat{\rho}\right)\nabla\hat{\chi}+\frac{U}{2}\hat{\rho}\left( \hat{\rho}-1\right)\right\}
\end{equation}
plus the necessary terms to enforce a fixed total particle number \cite{CHR06}. Observe that in the new variables, the
action is explicitly analytical.

We now split all variables into a homogeneous and an inhomogeneous part. 

\begin{equation}
\hat{\rho}\left( \mathbf{x},t\right) =n +\hat{r}\left( \mathbf{x},t\right) \label{decomp}
\end{equation}

\begin{equation}
\hat{r}\left( \mathbf{x},t\right)  =\sum_{\mathbf{p}\neq 0}\hat{r}_{\mathbf{p}}\left( t\right) f_{\mathbf{p}}\left( \mathbf{x}\right) 
\label{rp}
\end{equation}
where the $f_{\mathbf{p}}$ are plane waves

\begin{equation}
f_{\mathbf{p}}\left( \mathbf{x}\right) =\frac{1}{V^{1/2}}\exp\left\{i\mathbf{px}/\hbar\right\}
\end{equation}
and the allowed values of the components $p_{\mu}$, $\mu=1-3$, of the momentum $\mathbf{p}$ are integer multiples of $2\pi\hbar/L$,
and similarly

\begin{equation}
\hat{\chi}\left( \mathbf{x},t\right) =\frac{\hat{X}_0}{V^{1/2}}+\sum_{\mathbf{p}\neq 0}\hat{X}_{\mathbf{p}}\left( t\right) f_{\mathbf{p}}\left( \mathbf{x}\right) 
\label{xp}
\end{equation}
Observe that the homogeneous part of the density operator is constrained to be the c-number $n=N/V$, and the homogeneous part of the phase is a collective coordinate \cite{Raj87} which couples only to the homogeneous density. It does not affect the dynamics of the inhomogeneous modes.

Consider the lowest order theory which is obtained by keeping only the
 ``free'' quadratic part of the Hamiltonian 

\begin{equation}
\hat{H}_{\mathrm{free}}\left( \hat{r}_{\mathbf{p}},\hat{X}_{\mathbf{p}}\right)  =\sum_{\mathbf{p}\neq 0}\left\{\frac{\nu_p}{2}\left(n\hat{X}_{-\mathbf{p}}\hat{X}_{\mathbf{p}}-i\hat{r}_{-\mathbf{p}}\hat{X}_{\mathbf{p}}\right)+\frac{U}{2}\hat{r}_{-\mathbf{p}}\hat{r}_{\mathbf{p}}\right\}
\end{equation}
where $\nu_p=p^2/M$, $p=\left|\mathbf{p}\right|$. The Heisenberg equations of motion are 

\begin{equation}
\hbar\frac{d}{dt}\hat{X}_{\mathbf{p}}=\frac{-i\nu_p}{2}\hat{X}_{\mathbf{p}}+U\hat{r}_{\mathbf{p}}
\end{equation}

\begin{equation}
-\hbar\frac{d}{dt}\hat{r}_{\mathbf{p}}=n\nu_p\hat{X}_{\mathbf{p}}+\frac{-i\nu_p}{2}\hat{r}_{\mathbf{p}}
\end{equation}
Where

\begin{equation}
Un=\frac{4\pi\hbar^{2}Na}{ML^{3}}
\end{equation}
In the Bose Nova scenario, we have $U=0$ if $t\leq 0$. Therefore the frequencies are $\hbar\omega^{<} _{p}={\nu _p}/{2}$. If we call $A_{\mathbf{p}}$ the destruction operator which kills the initial state, then

\begin{equation}
r_{\mathbf{p}}\left( 0^{-}\right) =\left( -i\right) \sqrt{n}%
\left[ A _{\mathbf{p}}-A _{-\mathbf{p}}^{\dagger }\right]   
\end{equation}

\begin{equation}
X_{\mathbf{p}}\left( 0^{-}\right) =\frac{1}{\sqrt{n}} A_{\mathbf{p}} \label{xalpha0}
\end{equation}
For $t>0$, we have $U<0$ instead, and

\begin{eqnarray}
X_{\mathbf{p}} \left( t \right) &=& \frac{1}{{\sqrt n }}\left\{\left[\cos\left[ {\omega _p t} \right]-i\left(  \frac{\nu _p  }{2}+Un \right)\frac{\sin\left[ \omega _p t \right]}{\hbar\omega _p }\right]A_{\mathbf{p}}\right.\nonumber\\ &+&\left. iUn\frac{\sin\left[ \omega _p t \right]}{\hbar\omega _p }A_{ -\mathbf{p}} ^ {\dagger}\right\}\label{maineq}
\end{eqnarray}
with the dispersion relation

\begin{equation}
\omega _p=\frac{1}{\hbar}\sqrt{\nu _p\left( Un+\frac{\nu _p}{4}\right) }
\label{phonspec}
\end{equation}

\section{The one-body density matrix}

We may now turn to computing the one-body density matrix 

\begin{equation}
\sigma\left(\mathbf{x,y},t\right)=\left\langle \hat\Psi^{\dagger }\left( \mathbf{x},t\right) \hat\Psi\left(
\mathbf{y},t\right) \right\rangle \equiv \left\langle \exp i\left[ \chi
^{2*}\left( \mathbf{x},t\right)-\chi^{1}\left(
\mathbf{y},t\right)\right]  \right\rangle
\end{equation}
In the last term, the $1,2$ superindex indicates closed-time-path ordering: operators with a $2$ superindex always go to the left of operators with a $1$ superindex.
Observe that in our variables, the observable to be computed is a pure
exponential: there are no square roots to be developed. This is the whole
point of introducing the new variables.

As in the previous Section, we separate the variables $\chi^{2*}$ and $\chi^{1}$ into their homogeneous and inhomogeneous parts. Observe that the homogeneous terms may affect the overall normalization of the one particle density functional but not its shape. The overall normalization, on the other hand, is determined by the requirement that $\sigma\left(\mathbf{x,x},t\right)=n$. So we may simply continue to disregard the homogeneous terms. 

Since we have restricted ourselves to a Hamiltonian which is quadratic in the inhomogeneous modes, we may use the Wick theorem result

\begin{equation}
\left\langle e^{iA}\right\rangle =\left\langle 1\right\rangle\mathrm{Exp}\left\{\frac{-1}{2}\left\langle A^{2}\right\rangle\right\}
\end{equation}
with
\begin{equation}
A=A\left(\mathbf{x,y}\right)=\chi
^{2*}\left( \mathbf{x},t\right)-\chi^{1}\left(
\mathbf{y},t\right)
\end{equation}
Decomposing the field operators in modes, with due attention to the closed time path ordering, we obtain

\be
\left\langle A^{2}\right\rangle =\mathrm{const.}+2\sum_{\mathbf{p}\neq 0}\left[\frac 1V-f_{\mathbf{-p}}\left( \mathbf{x}\right)f_{\mathbf{p}}\left( \mathbf{y}\right)  \right]\left\langle X_{\mathbf{p}}^{\dagger}X_{\mathbf{p}}\right\rangle\label{ourresult}
\ee
Using the decomposition \eqn{maineq}

\begin{equation}
\left\langle A^{2}\right\rangle =\mathrm{const.}+\frac{4\left(Un\right)^2}{N}\sum_{\mathbf{p}\neq 0}\left\{\sin\left[\frac{ {\mathbf{p\left(x-y\right)}}}{2\hbar}\right]\right\}^{2} \left[\frac{\sin\left[ \omega _p t \right]}{\hbar\omega _p }\right] ^{2}
\end{equation}
To continue, we consider only the contribution from the unstable modes. The condition for instability is $U<0$ with $\left|Un\right|>\nu_p/4$. 
Since the lowest possible nontrivial value of $p$ is $h/L$, we get the critical scattering length as $a_c=\pi L/4N$. For $a$ close enough to the critical value, the six modes with $L^2p^2=h^2$ are the only unstable ones. Their frequency is $\omega =-i\gamma$, where $\gamma$ is given in \eqn{gamma} above. Setting $\mathbf{y}=0$ and $x=\left|\mathbf{x}\right|$, we get

\begin{equation}
\sum_{p=h/L}\left\{\sin\left[\frac{ \mathbf{px}}{2\hbar}\right]\right\}^{2} \left[\frac{\sin\left[ \omega _p t \right]}{\hbar\omega _p }\right] ^{2}\sim 2\left[\frac{\pi\sinh\left[ \gamma t \right]}{\hbar\gamma }\right]^2\left(\frac xL\right)^2
\end{equation}
therefore
\begin{equation}
\sigma \left(\mathbf{x},t\right)=n\:\mathrm{Exp}\left\{-\left[\frac{2\pi Un\sinh\left[ \gamma t \right]}{N^{1/2}\hbar\gamma L} \right]^{2}x^2\right\}
\end{equation}
The condensate occupation number $N_c$ is obtained by integrating over $\mathbf{x}$, so, once the Gaussian approximation becomes valid

\begin{equation}
N_c\propto\:e^{-3\gamma t}
\end{equation}
We therefore obtain the same scaling law as in \cite{CH03}, but the coefficient is $3/2$ times larger. As noted in the Introduction, this correction is enough to account for the anomaly observed in \cite{WDBDBHS06}.

\section{Comparison to other approaches}
In this Section we will compare the result above for the one-particle density matrix with other approaches in the literature, namely the particle-number conserving (PNC) formalism and the Hartree-Fock-Bogoliubov (HFB) and Popov approximation. We shall not discuss the so-called truncated Wigner approximation, but refer the reader to the detailed treatment in \cite{WDBDBHS06}. See \cite{CH08} for further details on these approaches.

\subsection{The equations of motion in the PNC approach}
The PNC formalism \cite{GirArn59,GirArn98,Gar97,CasDum98,GarMor06} is usually presented as an expansion in  inverse
powers of the total particle number $N.$ In preparation for this, it is convenient to scale
the interaction term, writing $U=u/N.$

The basic insight of the PNC approach is that if the total particle number remains constant, then each
particle above the condensate corresponds to a hole in the
condensate, so we may speak of \textit{particle-hole} (PH) pairs. 

Let us consider the expansion of the field operator in plane waves
\be \Psi\left( {\mathbf x},t\right) =\sum_{\mathbf{p}}a_{\mathbf{p}}\left( t\right) f_{\mathbf{p}}\left( \mathbf{x}\right)  \ee
$a_0$ reduces the number of particles in the condensate by one.
Following Arnowitt and Girardeau, let us introduce the operator

\be \beta =\frac 1{\sqrt{\hat N_0+1}}a_0=a_0\frac 1{\sqrt{\hat N_0}}
\ee where \be \hat N_0=N-\sum_{\mathbf{p} \neq 0}a_\mathbf{p} ^{\dagger
}a_\mathbf{p} \ee is the condensate number Heisenberg operator. Observe
that for a number eigenstate $\beta \left| N_0\right\rangle =\left|
N_0-1\right\rangle $
unless $N_0=0,$ in which case $\beta \left| 0\right\rangle =0.$ Therefore $%
\beta $ preserves the norm for all states orthogonal to the state
with no particles in the zeroth mode (which is much stronger than
not having a condensate). If there is a condensate, any physically
meaningful state will satisfy this requirement, and $\beta $ may be
considered a unitary operator, with inverse

\be \beta ^{\dagger }=\frac 1{\sqrt{\hat N_0}}a_0^{\dagger
}=a_0^{\dagger }\frac 1{\sqrt{\hat N_0+1}} \ee

We now introduce the destruction operator of a PH with the particle in mode $%
\mathbf{p} $ \be \lambda _\mathbf{p} =\beta ^{\dagger }a_\mathbf{p}. \ee If we
consider the $\beta $'s as unitary, then the $\lambda $'s satisfy
bosonic canonical commutation relations. This relationship may be
inverted

\be a_\mathbf{p} =\beta \lambda _\mathbf{p} \ee The number of particles in a
given mode is equal to the number of PH

\be a_\mathbf{p} ^{\dagger }a_\mathbf{p} =\lambda _\mathbf{p} ^{\dagger }\lambda
_\mathbf{p} \ee

We write the field operator restricted to the subspace with a well
defined total number of particles $N$ as $\Psi =\sqrt{N}\beta \phi $

\be \phi =\phi _0+\frac 1{\sqrt{N}}\lambda \left(
{\mathbf x},t\right) -\frac 1{2N}F\left[ \delta n\left( t\right) \right] \phi
_0\label{PNCexpan} \ee where for a homogeneous condensate we must have $\phi_0=V^{-1/2}$

\be \lambda \left( {\mathbf x},t\right) =\sum_{\mathbf{p}\neq 0}\lambda_{\mathbf{p}}\left( t\right) f_{\mathbf{p}}\left( \mathbf{x}\right)  \ee

\be \delta n\left( t\right) =\int d^{3}{\mathbf {x}}\;\lambda ^{\dagger }\lambda
\ee

\be F\left( x\right) =2N\left[ 1-\sqrt{1-\frac xN}\right] \sim
x+O\left( N^{-1}\right) \ee Within our approximations $\beta $
commutes with $\phi.$ To lowest order in $N^{-1}$, $\lambda$ evolves according to

\begin{equation}
0 = -i\hbar \lambda _{,t}+H \lambda
+Un\left( \lambda +\lambda ^{\dagger
}\right)
 +O\left( N^{-1/2}\right)\label{NLO}
\end{equation}
(see Appendix)

\subsection{The HFB and Popov approximations}
Before proceeding to compute the one particle density matrix in the PNC approach, let us show that the HFB and Popov approximations give essentially equivalent results.

The HFB and Popov approximations are implementations of the symmetry breaking approach to condensation, where the formation of a BEC is associated to the spontaneous breaking of the $U\left(1\right)$ symmetry \eqn{global} \cite{LEG06}. The field operator develops a c-number expectation value, which by traslation symmetry may depend only on time 

\be
\left\langle \Psi \right\rangle=e^{-i\Theta\left(t\right)}\Phi\left(t\right)
\ee
More generally

\be
\Psi =e^{-i\Theta\left(t\right)}\left[\Phi\left(t\right) +\psi\right]
\ee
In the HFB approach, we use this decomposition in the Heisenberg equations of motion, where we also replace products of two fluctuation operators by their expectation value, and use the so-called Hartree approximation. 

\be
\psi^{\dagger}\psi^2\sim 2\tilde{n}\psi+\tilde{m}\psi^{\dagger}
\ee
where

\be
\tilde{n}=\left\langle \psi^{\dagger}\psi\right\rangle
\ee

\be
\tilde{m}=\left\langle \psi^2\right\rangle
\ee
The Heisenberg equations decompose into equations for the mean fields and equations for the fluctuations

\be
i\hbar\frac{d}{dt}\Phi+\eta\Phi = U\Phi^3+2U\tilde{n}\Phi +U\tilde{m}\Phi
\ee

\be
i\hbar\frac{\partial}{\partial t}\psi+\eta\psi = H\psi +2U\left(\Phi^2+\tilde{n}\right)\psi +U\left(\Phi^2+\tilde{m}\right)\psi^{\dagger}
\ee
where

\be
\eta =\hbar\frac{d\Theta}{dt}
\ee
The HFB approximation has the serious drawback that it is not gapless, and therefore it is hardly reliable in a problem such as the Bose Nova, which depends critically on the behavior of long wavelength modes. The Popov approximation overcomes this problem by further neglecting $\tilde{m}$. Then we obtain

\be
\eta =U\Phi^2+2U\tilde{n}
\ee
and the fluctuation equation becomes
\be
i\hbar\frac{\partial}{\partial t}\psi = H\psi +U\Phi^2\left(\psi +\psi^{\dagger}\right)
\ee
Under this approximation the number of particles in the condensate remains constant. This may be avoided by including explicitly the effect of particle loss through three body recombination. However, the final results are robust against these terms \cite{SU03,ADH04,WDBDBHS06}, and we shall not consider them in detail. On the other hand, the total number of particles is not conserved.

If we assume that the temperature is effectively absolute zero, then $\Phi^2=n$ initially and remains close to it until much later in the collapse; the effect of finite temperature is discussed in \cite{WDBDBHS06} and it is seen to be minor. If we just replace $\Phi^2=n$, the Popov equation for the fluctuations reduces to the PNC equation for the inhomogeneous modes \eqn{NLO}. This approximation gives a reasonable description of early jet and burst formation \cite{CH03}, so it may be considered reliable.

\subsection{The one particle density matrix in the PNC approach}
We now return to the calculation of the one particle reduced density matrix. 

\begin{eqnarray}
\sigma\left(\mathbf{x,y},t\right)&=&\left\langle \hat\Psi^{\dagger }\left( \mathbf{x},t\right) \hat\Psi\left(
\mathbf{y},t\right) \right\rangle \nonumber\\&\equiv& n\left\{1-\frac 1N\left[\left\langle \delta n\right\rangle-V\left\langle \lambda^{\dagger}\left( \mathbf{x},t\right)\lambda\left(
\mathbf{y},t\right)\right\rangle\right]\right\}
\end{eqnarray}
Decomposing in modes, we get

\be
\sigma\left(\mathbf{x,y},t\right)= n\left\{1-\frac 1n\sum_{\mathbf{p}\neq 0}\left[\frac 1V-f_{\mathbf{-p}}\left( \mathbf{x}\right)f_{\mathbf{p}}\left( \mathbf{y}\right)  \right]\left\langle \lambda_{\mathbf{p}}^{\dagger}\lambda_{\mathbf{p}}\right\rangle\right\}\label{PNCresult}
\ee
Each mode evolves according to

\be
i\hbar\frac{d\lambda_{\mathbf{p}}}{dt}=\frac{\nu_p}2\lambda_{\mathbf{p}}+Un\left(\lambda_{\mathbf{p}}+\lambda_{-\mathbf{p}}^{\dagger}\right)
\ee
The dispersion relation is given by \eqn{phonspec}. At $t=0$, $\lambda$ must destroy the physical state, so $\lambda_{\mathbf{p}}\left(0\right)=e^{i\varphi_{\mathbf{p}}}A_{\mathbf{p}}$ for some phase $\varphi_{\mathbf{p}}$. From the equation of motion we derive the initial velocity

\be
i\hbar\frac{d\lambda_{\mathbf{p}}}{dt}\left(0\right)=\frac{\nu_p}2e^{i\varphi_{\mathbf{p}}}A_{\mathbf{p}}+Un\left(e^{i\varphi_{\mathbf{p}}}A_{\mathbf{p}}+e^{-i\varphi_{\mathbf{p}}}A_{-\mathbf{p}}^{\dagger}\right)
\ee
Therefore

\begin{eqnarray}
\lambda_{\mathbf{p}} \left( t \right) &=& \left\{\left[\cos\left[ {\omega _p t} \right]-i\left(  \frac{\nu _p  }{2}+Un \right)\frac{\sin\left[ \omega _p t \right]}{\hbar\omega _p }\right]e^{i\varphi_{\mathbf{p}}}A_{\mathbf{p}}\right.\nonumber\\ &-&\left. iUn\frac{\sin\left[ \omega _p t \right]}{\hbar\omega _p }e^{-i\varphi_{\mathbf{p}}}A_{ -\mathbf{p}} ^ {\dagger}\right\}\label{maineq2}
\end{eqnarray}
This equation and \eqn{maineq} show that

\be
\left\langle \lambda_{\mathbf{p}}^{\dagger}\lambda_{\mathbf{p}}\right\rangle =n\left\langle X_{\mathbf{p}}^{\dagger}X_{\mathbf{p}}\right\rangle
\ee
and therefore the PNC result \eqn{PNCresult} is just the first term in the expansion of our earlier result \eqn{ourresult} in inverse powers of $N^{1/2}$.

Indeed, the representations of the field operators \eqn{represen1} and \eqn{PNCexpan} are equivalent, to next to leading orden in $N^{-1/2}$, provided we identify $e^{i\varphi_{\mathbf{p}}}=-i$ and $\exp\left\{-i\hat{X}_0/V^{1/2}\right\}=n^{1/2}\beta$.

\section{Final remarks}
After this point, it only remains to comment on the reasons why this proposal works. 

From the formal point of view, our expression for the one particle reduced density matrix is seen to agree with the perturbative implementation of the particle number conserving approach to next to leading order. This agreement suggests that, more generally, our approach implements a resummation of the PNC expansion. A key feature is that the we use variables that keep the exponential structure of the one  particle density matrix. Therefore, the method suggested amounts to a perturbative evaluation of the exponent, but it is non perturbative with respect to the final result.

This formal advantage of the proposed method correlates with a shift in the physical emphasis, from particle creation in the excited modes to the loss of coherence among atoms. Comparing this to other formal studies of decoherence, it comes as no surprise that the later process is faster than the former \cite{ALMV06,LoRiVi07}.

We submit this minor contribution with the expectation that it will help clear the way to a full quantitative understanding of this fascinating experiment.

\section{Acknowledgments}
It is a pleasure to acknowledge many discussions with Bei-lok Hu.

This work is supported in part by University of Buenos Aires, CONICET and ANPCyT (Argentina)
\section*{Appendix: Derivation of \eqn{NLO}}
The idea is to seek a solution of the Heisenberg equations of motion for $%
\Psi $ where $\beta $ and the $\lambda $'s have developments
in inverse powers of $N$. Define a q-number chemical potential
$\hat \mu $ from

\be \beta ^{\dagger }\frac{d\beta }{dt}=\frac{-i\hat \mu }\hbar \ee

We have

\be i\hbar \frac \partial {\partial t}\phi =\left( H-\hat \mu
\right) \phi +u\phi ^{\dagger }\phi ^2 \ee

We then find

\begin{eqnarray}
0 &=&-\hat \mu \phi _0+u\phi _0^3 \nn
&&+\frac 1{\sqrt{N}}\left[ -i\hbar \lambda _{,t}+\left( H-\hat \mu
\right) \lambda +u\phi _0^2\left( 2\lambda +\lambda ^{\dagger
}\right) \right]\nn &&+ O\left( N^{-1}\right)
\end{eqnarray}
Taking the expectation value we find

\begin{eqnarray}
0 &=&-\left\langle \hat \mu
\right\rangle \phi _0+u\phi _0^3-\frac
1{\sqrt{N}}\left\langle \hat \mu \lambda
\right\rangle  \nn
&&\ +O\left( N^{-1}\right)
\end{eqnarray}
Recall that $\hat \mu $ is Hermitian. So we may decompose this equation into

\begin{equation}
0 =-\left\langle \hat \mu \right\rangle \phi
_0+u\phi _0^3-\frac 1{2\sqrt{N}}\left\langle \hat \mu \lambda
+\lambda ^{\dagger }\hat \mu \right\rangle
 +O\left( N^{-1}\right)\label{pncf1}
\end{equation}
and

\begin{equation}
0 =\frac 1{2\sqrt{N}}\left\langle \hat \mu
\lambda
-\lambda ^{\dagger }\hat \mu \right\rangle
+O\left( N^{-1}\right)\label{pncf2}
\end{equation}
Subtracting the expectation value from the Heisenberg equation, we
get

\begin{eqnarray}
0 &=&\left( \left\langle \hat \mu \right\rangle -\hat \mu \right) \phi _0 \nn
&&\ +\frac 1{\sqrt{N}}\left[ -i\hbar \lambda _{,t}+\left( H-\hat \mu
\right) \lambda +u\phi _0^2\left( 2\lambda +\lambda ^{\dagger
}\right) \right]
+\frac 1{\sqrt{N}}\left\langle \hat \mu \lambda \right\rangle \nn
&&\ +O\left( N^{-1}\right)\label{pncf3}
\end{eqnarray}
and from \eqn{pncf3},\eqn{pncf2} and \eqn{pncf1} we get

\begin{equation}
\hat \mu =\left\langle \hat \mu \right\rangle 
+O\left( N^{-1}\right)\sim\frac uV=Un
\end{equation}
Observe that this implies
\be
\left\langle \hat \mu\lambda \right\rangle =O\left( N^{-1/2}\right)
\ee
The equation for
$\lambda $ simplifies into \eqn{NLO}

\end{document}